# AI and Human Oversight: A Risk-Based Framework for Alignment

**Laxmiraju Kandikatla, MPharm, CQA**
MaxisIT Inc., Edison, NJ, United States
Aula Fellowship for AI, Montreal, Canada
laxmiraju.kandikatla@gmail.com

**Branislav Radeljić, PhD, SFHEA**
Aula Fellowship for AI, Montreal, Canada
branislav@theaulafellowship.org
ORCID: 0000-0002-0497-3470

**Abstract**
As Artificial Intelligence (AI) technologies continue to advance, protecting human autonomy and promoting ethical decision-making are essential to fostering trust and accountability. Human agency—the capacity of individuals to make informed decisions—should be actively preserved and reinforced by AI systems. This paper examines strategies for designing AI systems that uphold fundamental rights, strengthen human agency, and embed effective human oversight mechanisms. It discusses key oversight models, including Human-in-Command (HIC), Human-in-the-Loop (HITL), and Human-on-the-Loop (HOTL), and proposes a risk-based framework to guide the implementation of these mechanisms. By linking the level of AI model risk to the appropriate form of human oversight, the paper underscores the critical role of human involvement in the responsible deployment of AI, balancing technological innovation with the protection of individual values and rights. In doing so, it aims to ensure that AI technologies are used responsibly, safeguarding individual autonomy while maximizing societal benefits.

**Keywords**: AI systems, decision-making, human oversight, risk assessment, societal benefits

## 1. Introduction

Artificial Intelligence (AI) systems are playing an increasingly influential role across various sectors. Abilities once thought to be uniquely human—like intuition, moral judgment, and inventive thinking—are now being rivaled by machines capable not only of mimicking these traits but, in many specialized fields, surpassing them (Fui-Hoon Nah et al., 2023; Mikalef et al., 2022). As technological progress accelerates beyond the capacity of most individuals to keep up, artificial intelligence is increasingly taking center stage in reshaping how we define labor, efficiency, and even meaning itself. In this emerging landscape, AI is frequently entrusted with organizational structures, role assignment, and productivity fine-tuning (Budhwar et al., 2022; Mackenzie et al., 2025). Accordingly, some studies insist that "[d]eveloping machine morality is crucial, as we see more autonomous machines integrated into our daily lives" (Chu & Liu, 2023, also, Jiang et al., 2021). This is even more relevant in light of the argument maintaining that "while AI can improve efficiency, it may also reduce critical engagement, particularly in routine or lower-stakes tasks in which users simply rely on AI" (Bois, 2025; also, Schmidt et al., 2020; Vasconcelos et al., 2023).

The interaction between humans and machines reflects deeper existential concerns about power relations and human agency in the problem-solving process, especially given that inadequate oversight or failure to identify model errors can lead to serious and potentially irreversible harm. For example, healthcare and clinical research are particularly delicate areas due to their direct impact on patient safety, making careful implementation essential (Elgin & Elgin, 2024; Rodrigues, 2000). Finance is highly sensitive as it directly impacts individual wealth, market stability, and fraud risks, requiring robust governance and risk controls (Moura et al., 2025). Manufacturing carries significant risk since errors can disrupt supply chains, compromise product quality, and create safety hazards, necessitating precise implementation and continuous monitoring (Aljohani, 2023; Khurram et al., 2025). Similarly, education is a critical sector because of its influence on student learning outcomes and future opportunities, demanding thoughtful and equitable deployment of technology. Transportation is equally vital, as system failures can result in accidents and loss of life, making reliable, safe, and well-regulated deployment indispensable.

All of this points to the need for robust governance, which should not in itself be a major issue, since policymakers, whose engagement is pivotal in shaping the future, are quick to argue that their approaches and regulations are informed by a range of factors, including past experiences and risk management strategies (Law & McCall, 2024; Perry & Uuk, 2019; Wirtz et al., 2021). The process is further complicated by AI systems, which rely heavily on programming languages and coded algorithms; this suggests that responsibility for their actions extends to programmers, data scientists, content providers, institutional architects, and many others, all of whom contribute to shaping their design, training, and deployment (Gebru & Torres, 2024; Gerlich, 2025; Rafanelli, 2022). However, since human agency entails the capacity to make autonomous and informed decisions, the implementation of AI systems should support, not replace, human decision-making; in other words, rather than focusing on how AI could or will replace humans, it is more appropriate to emphasize the degree of human oversight required (Langer et al., 2025; Sterz et al., 2024; Sturgeon et al., 2025). This question of oversight is particularly important given the potential harm that AI development and deployment may pose to individual rights. This concern underpins the argument that ethical AI design should prioritize human agency while leveraging the efficiencies of AI-driven assistance.

This paper advances the conversation on trustworthy AI (EU, 2019) by offering a structured approach to embedding human oversight models—Human-in-Command (HIC), Human-in-the-Loop (HITL), and Human-on-the-Loop (HOTL)—into risk assessment and mitigation strategies for AI systems. The paper argues that responsible AI deployment requires more than technical robustness; it also demands context-sensitive oversight aligned with the level of risk posed by the system's decisions. We first propose a practical framework for scenario identification, in which AI use cases are classified according to their potential impact on human well-being, safety, and compliance obligations. Building on this classification, we apply a risk assessment methodology to determine where human involvement is critical and what level of oversight is required. The paper puts forward a set of guiding principles: (1) HIC ensures governance-level accountability, where humans retain ultimate authority over system objectives, as in public policy or defense applications; (2) HITL mechanisms are crucial when real-time human judgment can prevent harm, such as in clinical decision support systems; and (3) HOTL is suited for systems where monitoring and periodic intervention can mitigate risks without slowing operations, such as financial fraud detection. To demonstrate the proposed framework, we have considered examples from multiple domains, including finance (AI/ML-based prediction for bank loan approvals; see Haque & Hassan, 2024; Sheikh et al., 2024; Uddin et al., 2023), education (AI and student performance prediction; see Ahmed et al., 2025; Hoti et al., 2025; Johora et al., 2025; Karale et al., 2022; Tirumanadham et al., 2025), and healthcare (automated patient scheduling using AI; see Algarvio, 2025; Knight et al., 2023; Schmider, 2019). Using these examples from the relevant literature, we performed a risk assessment based on our proposed framework and applied corresponding human oversight mechanisms. In addition to these oversight models, appropriate control measures—technical, procedural, and organizational—are recommended to operationalize trust and ensure compliance. In doing so, the paper contributes a repeatable, risk-based process that organizations can adopt to ensure AI systems remain safe, ethical, and aligned with human values.

## 2. The state of the debate

Artificial Intelligence has emerged as one of the most pivotal technologies of the 21st century, with rapid advancement that is transforming industries, reshaping the global economy, and influencing nearly every aspect of daily life (Bellaiche et al., 2023; Broekhuizen et al., 2023; Hutson & Rains, 2025; Sahebi & Formosa, 2025; Smuha, 2025). However, AI standards are not inherently objective or impartial; rather, they are shaped by institutional priorities, political motivations, and value-laden decisions rooted in particular regional and cultural contexts (Dubber et al., 2021; Robinson, 2025; Solow-Niederman, 2023; Tallberg et al., 2023; Zekos, 2022). This becomes especially significant when AI technologies are used to reinforce or legitimize existing hierarchies and power dynamics; in other words, AI is capable of masking ideological bias, reinforcing inequality, and undermining democratic deliberation (Arora et al., 2023; Gardiner, 2024; Kleinberg et al., 2018). Moreover, as has been aptly cautioned, AI platforms possess the ability to impact national security and shape public discourse in ways that often evade democratic oversight; in doing so, they can inadvertently intensify social exclusion and enable pervasive surveillance, placing strategic or geopolitical interests above the broader public good (Le Mens & Gallego, 2025).

Regarding human oversight, there is a general agreement on its importance not only to ensure accountability but also to support improvement for the benefit of all (Binns, 2022; Crootof et al., 2023; Solove & Matsumi, 2024; Wagner, 2019). For example, in addition to advocating for a "strict regulatory oversight," Article 14 of the European Union AI Act stipulates that "[h]uman oversight shall aim to prevent or minimize the risks to health, safety or fundamental rights that may emerge when a high-risk AI system is used in accordance with its intended purpose or under conditions of reasonably foreseeable misuse" (EU, 2024; also, Laux, 2024; Tomić & Štimac, 2025). Such a remark suggests that while AI systems are driven by huge pools of data and algorithms that power decision-making processes, they risk being unable to recognize, let alone prioritize, ethical considerations (Enarsson et al., 2022; Kordzadeh & Ghasemaghaei, 2022; Salloch & Eriksen, 2024). In addition, and looking from a different perspective, others point out that since "people are unable to perform the desired oversight functions," which means that "human oversight policies legitimize government uses of faulty and controversial algorithms without addressing the fundamental issues with these tools," it would be more appropriate to consider the prospect for "a shift from human oversight to institutional oversight as the central mechanism" (Green, 2022).

Linked to the above is the question of training, including fine-tuning and self-replication. Put simply, every sector should be able to identify and provide the most effective training, enabling those in charge to recognize potential flaws and verify facts and critical information; this helps to prevent unintentional breaches, reputational damage, and potential profit loss (de Almeida & dos Santos, 2025). In light of this and given the nuances that characterize AI models and technological advancement more broadly, while those responsible for human oversight should undergo rigorous training and regular reviews, it is equally important that they be entrusted with sufficient authority to carry out their duties in accordance with established standards (Floridi et al., 2018). In this context, authority implies the independence to question, challenge, and critically assess AI outputs—not to defer to them as infallible experts, but to treat them as tools subject to human judgment and accountability. However, this discussion should not be isolated from the interests of corporate elites and trends in global capitalism, nor from the argument that the AI race risks prioritizing the agendas of powerful nations over responsible innovation (Bernstein, 2024; Smuha, 2021; Sublime, 2023; Thornhill, 2024).

Thus, who should be responsible for promoting oversight, especially given recent analyses highlighting the growing alignment between political and corporate elites, as well as the remarkable power of the largest tech companies, which remain detached from issues like poverty, unemployment, and economic uncertainty (Graham & Warren, 2023; McCourt, 2024; Schaake, 2024)? The emergence of AI ethics boards, independent AI audits, and algorithmic impact evaluations points to a growing emphasis on accountability. Yet, this trend is marked by certain inconsistencies. As rightly observed, "the understanding of how [the principles of responsible AI] can be operationalized in designing, executing, monitoring, and evaluating AI applications is limited" (Papagiannidis et al., 2025; also, Corrêa et al., 2023; Radanliev, 2025; Schuett et al., 2025). In any case, despite public commitments to ethical values, many corporations still place greater importance on profit and competitive advantage. In some instances, the promotion of ethical AI serves more as a marketing tool than a genuinely enforced standard, prompting doubt and criticism from civil society organizations and human rights advocates (Radeljić, 2024). In other words, given the blurred boundaries between public interest and private gain, and the increasing fragility of democratic systems under the weight of surveillance capitalism, data monopolies, and digital authoritarianism, it is reasonable to be skeptical not only about human involvement but also the level of oversight needed—HIC, HITL, and HOTL roles. While human intervention may be glorified, the space for its active performance, including the ability to override AI decisions, may be limited.

Aware of the overall complexity, and the fact that the degree of human oversight runs the risk of being closely tied to varying and potentially controversial interests, but also costs and the type of political system in place (ranging from democratic to authoritarian, or even totalitarian), it is more appropriate to link the question of oversight to sector-specific requirements, taking into account both safety standards and desired policy outcomes. To fill this gap, this paper proposes a structured and comprehensive framework for integrating human oversight into AI governance. Rather than focusing solely on abstract ethical principles, the paper associates oversight models Human-in-Command (HIC), Human-in-the-Loop (HITL), and Human-on-the-Loop (HOTL) to each sector, explicitly considering safety standards, compliance obligations, and desired policy outcomes. We extend the debate by moving beyond conceptual discussions of "human oversight" toward an actionable methodology (Figure 1): (1) identifying AI scenarios based on their impact on human well-being, safety, and regulatory context; (2) performing a risk assessment to determine the risk of AI Systems; and (3) mapping risk determinations to different levels of human oversight mechanisms. In doing so, this paper offers a replicable decision-making process that acknowledges contextual differences in healthcare, finance, education, or critical infrastructure while still promoting consistency in risk mitigation. This contribution ensures that human oversight should not be arbitrary or politically driven but should align with AI usage and societal values.

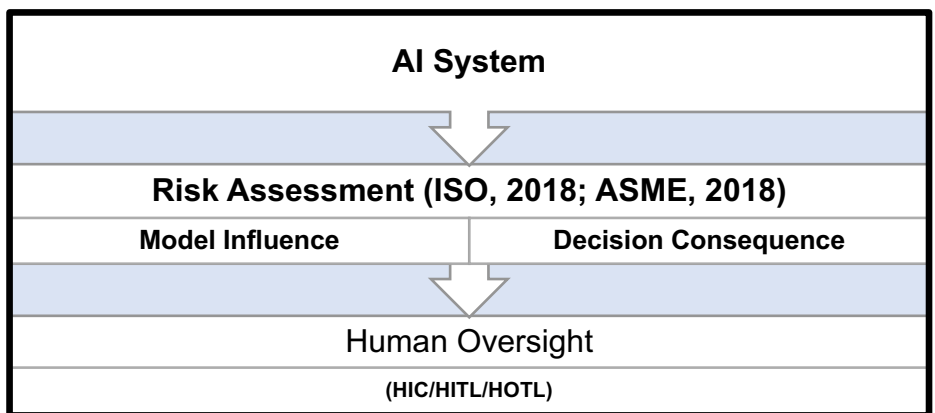

**Figure 1.** Risk assessment framework (Authors' proposal)

## 3. Human agency and oversight requirements
### 3.1. Human agency in AI

The deployment of AI raises concerns about a lack of fairness, transparency, and accountability, which could potentially infringe upon individual rights and freedoms (Nikolinakos, 2023). Thus, it is essential to conduct a Fundamental Rights Impact Assessment (FRIA) before the deployment of any high-risk AI systems (Figure 2) to ensure that human rights are safeguarded (Mennella et al., 2024; also, Mantelero, 2024). A FRIA is a structured process to evaluate how an AI system may affect human rights before deployment, especially for high-risk applications. While promoting transparency, responsible use, and public trust in AI, FRIA helps to identify and mitigate risks to rights such as privacy, non-discrimination, freedom of expression, and due process. The recently published EU Artificial Intelligence Act establishes a requirement to conduct a FRIA (EU, 2024); in other words, it mandates the assessment of risks posed by AI systems to health, safety, and fundamental rights. To address this challenge and facilitate compliance with the EU's AI Act Article 27, Pandit and Rintamäki (2024) present a novel representation of the FRIA as an ontology, utilizing W3C Semantic Web standards (Resource Description Framework (RDF), RDF Schema, and Web Ontology Language (OWL)) to create a structured, machine-readable, and interoperable model.

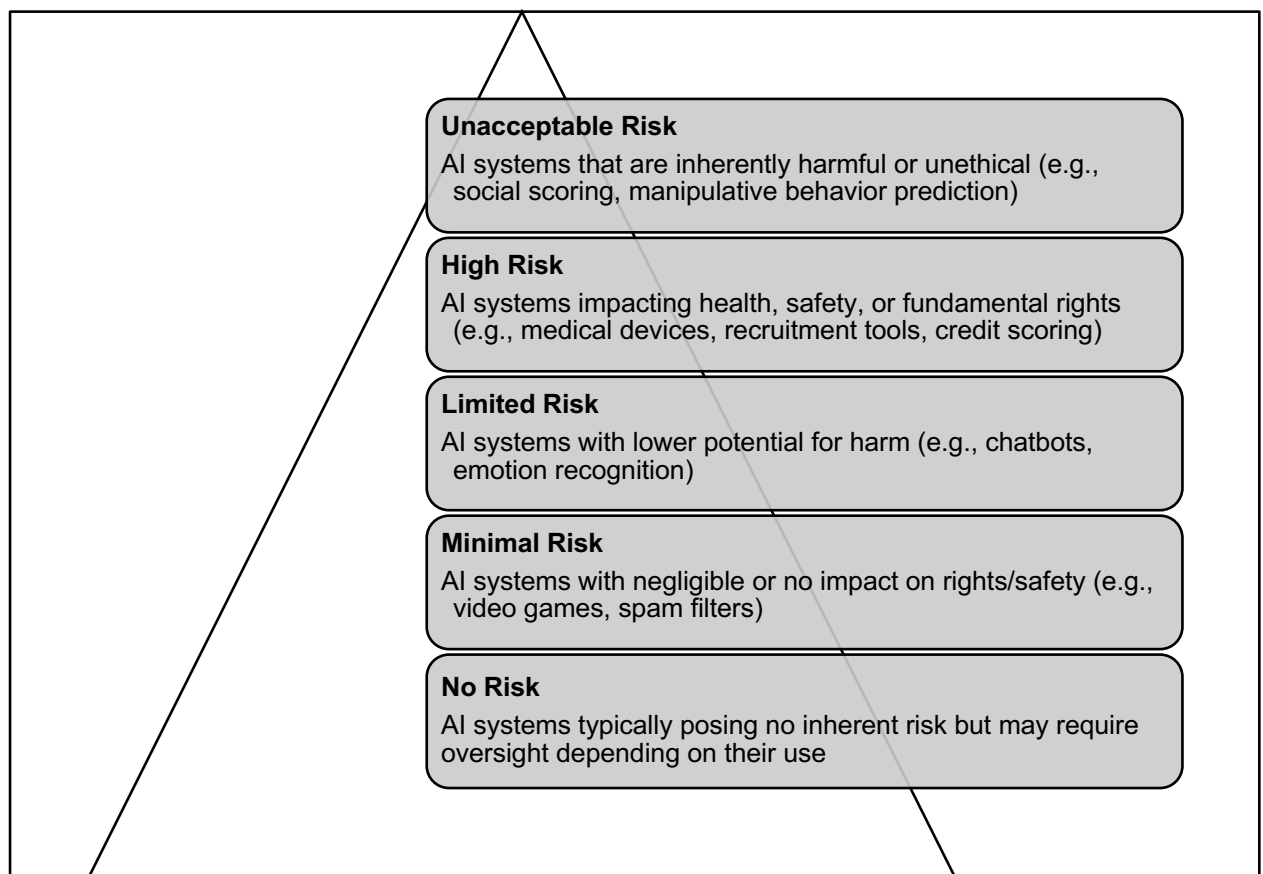

**Figure 2.** AI categories under the EU AI Act (adapted from Mennella et al., 2024)

Preserving human agency in AI involves designing systems that empower individuals to make autonomous and informed choices. AI systems should offer insights, recommendations, and data-driven analyses that support human judgment (Lu et al., 2024). This principle becomes particularly crucial in sectors like healthcare and clinical research, where decisions have significant consequences on patient safety. An example of this is an AI system used to detect diabetic retinopathy through retinal screening. While the AI model can accurately identify potential cases, it is recommended that the physician review its findings to avoid false positives, taking into account the patient's reports and physical examination (Schmider et al., 2019). This approach reinforces human agency and supports holistic care. Ethical AI in healthcare should prioritize human oversight, enabling professionals to make informed decisions and improve patient outcomes. Human oversight mechanisms are further explained in section below.

**3.2. Human oversight (HIC, HITL, and HOTL)**
While AI excels at processing large volumes of data rapidly, it often lacks an understanding of nuanced human values and experiences. Without oversight, AI may reinforce biases, misalign with user needs, or encourage blind adherence to algorithmic decisions; therefore, human oversight safeguards ethical use, upholds human rights, and maintains accountability. In high-stakes domains like healthcare or criminal justice, AI should support, not replace professional judgment because these areas involve ethical complexity, accountability, and human dignity, which may not be fully captured or managed by algorithms (Alowais et al., 2023).

Human-in-Command (HIC) places humans as the ultimate authority over AI systems, even in high-risk or autonomous settings (EU, 2019). In HIC, humans retain control, especially in sensitive or critical situations. For example, An AI system predicts potential safety risks in a clinical trial based on patient data (Ferrara et al., 2024). It is suggested that a researcher reviews the AI's predictions, considering the overall risk-benefit profile, data integrity, and clinical relevance. Then the researcher to approve, modifies, or rejects the AI's recommendations to ensure patient safety and trial integrity. HIC ensures humans have the final say and they are in command of these operations, even when AI performs autonomously.

Human-in-the-Loop (HITL) refers to active human involvement in AI decision-making processes, where humans provide real-time feedback to guide or correct the system's outputs (EU, 2019). This approach is essential when AI systems operate in high- or medium-risk contexts and may have significant impact if the system cannot make reliable, ethical, or safe decisions independently. For example, an AI system screens potential participants for a clinical trial based on eligibility criteria (Schmider et al., 2019). It is suggested that a researcher reviews the AI's selections to confirm the AI's decisions or based on risk indicators considering the patient's overall health, medical history, and potential interactions with other medications. HITL frameworks are designed to ensure that AI systems augment, not replace, human expertise by integrating human oversight into AI decision-making processes (EU, 2019). By actively involving humans in decision-making processes, organizations can leverage the benefits of AI while maintaining accountability and ethical standards.

Finally, Human-on-the-Loop (HOTL) refers to human supervision of AI systems with low risk. While the AI operates independently, humans monitor performance and intervene as needed or when anomalies occur (EU, 2019). For example, An AI system is used to automate the case processing of adverse events (non-serious) in clinical trials (Caixinha Algarvio et al., 2025), based on the patient records. In this scenario, if the AI model autonomously processes most non-serious cases and flags those involving issues it cannot handle—such as coding discrepancies, concerns about seriousness or causality, or an inability to interpret the nature of the adverse event—human intervention is required to resolve these complexities. HOTL frameworks strike a balance between autonomy and oversight by enabling efficient AI operations while preserving human intervention capabilities for critical decisions.

In any case, human oversight mechanisms necessitate that a human expert reviews, validates, or potentially overrides AI decisions. While this process is essential for ensuring accuracy and accountability, it can also result in delays. In time-sensitive scenarios, such as emergency medical situations or urgent decisions, this added layer of complexity may hinder swift decision making. However, it is crucial to implement oversight mechanisms that are proportionate to the level of risk and the impact of the decision. For example, in the medical field, the participation of patients as "co-reasoners" or "fellow-workers," suggesting that their participation can counter automation bias, can prompt physicians to engage more critically with AI outputs, and improve contextual reasoning in settings such as melanoma detection (Herrmann et al., 2024). This supports the case for HITL approaches where patient engagement contributes to safer, more transparent decision-making. However, since patient contributions are primarily psychological rather than epistemic, and without structural best practices, HITL participation alone cannot ensure systematic oversight. Extending this discussion further, other studies caution that integrating humans into the loop can inadvertently create a "moral crumple zone," where both clinicians and patients are scapegoated for failures rooted in opaque AI behavior (Ranisch et al., 2024). This risk of misattributed blame underscores the need for clear responsibility frameworks. Merely inserting humans into the decision cycle does not resolve accountability challenges and may create new responsibility gaps. Together, these insights strengthen the case for a risk-based oversight model that explicitly links HIC, HITL, and HOTL to scenario-specific controls.

## 4. Risk assessment and AI oversight

Effective AI risk management necessitates early and continuous communication and consultation with all relevant stakeholders, including AI developers, end-users, regulators, and domain experts (ISO, 2018). Across industries, embedding human oversight at this stage ensures that diverse stakeholder perspectives inform risk identification and decision-making processes, thereby enhancing transparency and fostering public trust. By integrating human judgment from the outset, organizations can more accurately anticipate the potential influence of AI outputs and the ramifications of incorrect decisions (Pandit & Rintamäki, 2024). Defining the scope, context, and risk criteria of AI deployment is crucial for determining the potential impact of AI across different applications (ISO, 2018) (Figure 3). Human oversight guides the identification of high-risk scenarios where AI outputs have substantial influence on critical decisions (US FDA, 2025).

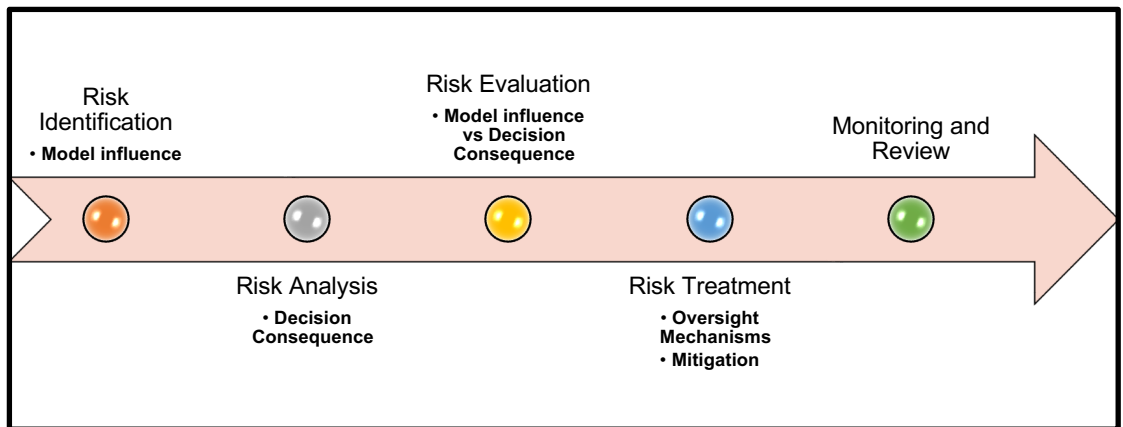

**Figure 3.** Risk assessment process mapping (adapted from ISO, 2018, and ASME, 2018)

### 4.1. Risk identification and analysis

AI system risk should be conceptualized not as an intrinsic attribute of the system itself, but as a risk that arises from deploying the system to address a specific question or decision-making context. As already suggested elsewhere (ASME, 2018; DiFrancesco, 2025; US FDA, 2025), this risk can be characterized along two principal dimensions: system influence and decision consequence. System influence pertains to the relative weight attributed to the AI system's outputs compared to other sources of evidence, such as clinical studies, operational data, expert judgment, or bench testing. Decision consequence addresses the potential impact of an erroneous decision based on AI output. A systematic evaluation of decision consequence must consider three elements: severity of potential harm, probability of occurrence, and detectability of an error prior to adverse outcomes. Assessing risk along these dimensions provides a structured framework that minimizes the risk of under- or overestimation.

### 4.2. Risk evaluation

Risk evaluation for AI systems involves assessing both model risk and decision consequence to provide a comprehensive understanding of potential adverse outcomes. Model risk is characterized by the degree of influence the AI system exerts on decision-making and the potential consequences of erroneous outputs (DiFrancesco, 2025; US FDA, 2025). For instance, AI systems with high influence where outputs form a critical basis for decisions such as clinical treatment plans or financial approvals carry elevated risk compared to systems where AI insights constitute only one factor among many. Decision consequence refers to the intended outcomes and impact the system's deployment aims to achieve, which can vary from low to high severity depending on the application context. The interplay of these factors defines risk levels ranging from low, low-medium, medium, medium-high, to high (Figure 4). Higher risk levels correspond to scenarios where both the influence of the AI system is significant and the consequences of wrong decisions are severe, necessitating more stringent risk mitigation and oversight. This framework supports targeted governance approaches that consider both the systemic role of AI and the contextual consequences of its use, enabling organizations to better prioritize resources and implement safeguards mechanisms effectively.

| | Risk Matrix | **Decision Consequence** | | |
|---|---|---|---|---|
| | | Low | Medium | High |
| **Model Influence** | **High** | Medium | Medium-High | High |
| | **Medium** | Low-Medium | Medium | Medium-High |
| | **Low** | Low | Low-Medium | Medium |

**Figure 4.** Risk assessment matrix (adapted from ASME, 2018)

### 4.3. Risk treatment

Following risk characterization, the level of human oversight should be mapped proportionally to the identified risk category. We are proposing below oversight mechanism based on the risk matrix (Figure 5). At the highest levels of risk, particularly within safety- or life-critical contexts, a HIC oversight is warranted. Humans retain primary authority over decision-making, and the AI system functions strictly as a decision-support tool rather than an autonomous actor. For medium to medium-high risk settings, a HITL framework is more suitable, ensuring that a human decision maker validates or approves the AI system's outputs before actions are executed. For low- to medium-risk applications, risks may be appropriately managed under a HOTL framework, whereby the AI system operates autonomously but remains under continuous human supervision, with intervention triggered in response to anomalies (EU, 2019; also, Chiodo et al., 2025; Fabiano, 2025; Holzinger et al., 2025; Tomić & Štimac, 2025).

This risk-to-oversight mapping is relevant across multiple sectors. In finance, fraud detection systems may function acceptably with HOTL monitoring, whereas credit approval decisions typically require HITL review. In manufacturing, predictive maintenance systems may operate autonomously with HOTL oversight, while production shutdown decisions necessitate HITL or HIC control due to their operational consequences. In autonomous vehicles, everyday navigation may be automated, but safety-critical maneuvers such as collision avoidance require HITL confirmation or strict HIC authority. Aligning human oversight intensity with the severity, probability, and detectability of system-related risks ensures that AI deployments achieve both efficiency and safety across diverse domains.

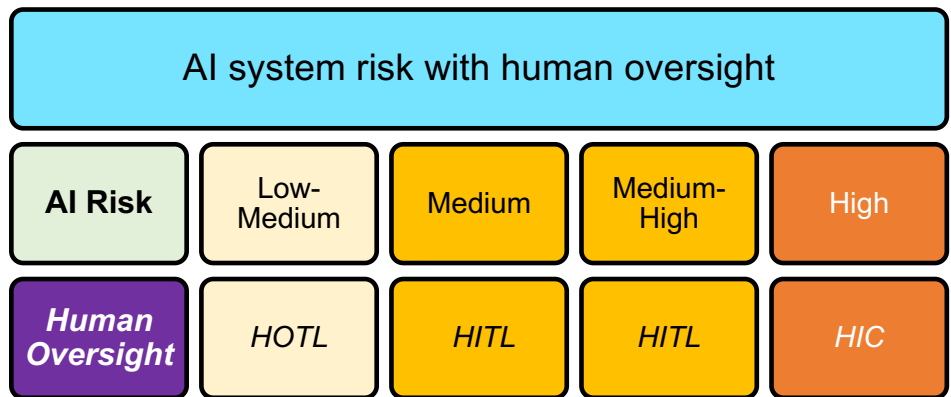

**Figure 5.** Human oversight mechanisms based on risk (Authors' proposal)

### 4.4. Risk monitoring and review

The scope of risk monitoring and review is to ensure that mitigation strategies are effectively implemented and that AI systems remain reliable over time (Table 1). Oversight mechanisms validate AI outputs, embed fail-safes for high-consequence decisions, and establish governance protocols to clarify accountability. Continuous review, recording, and reporting enable dynamic adjustments of oversight intensity as risk assessments evolve. Across industries, this approach ensures that AI systems operate safely, human judgment complements automation appropriately, and accountability is maintained, balancing efficiency with ethical and operational responsibility (ISO, 2018; also, Madiega, 2021; Pandit & Rintamäki, 2024). Therefore, it is recommended that both model influence and decision consequence be systematically evaluated to determine the appropriate level of oversight for AI systems.

| ISO 31000 Step | AI-Specific Concept | Description |
|---|---|---|
| Risk Identification | Model Influence | Extent to which AI output shapes outcomes (high influence = central role in decision) and identifies potential points of failure or misuse |
| Risk Analysis | Decision Consequence | Evaluate the potential impact of an incorrect AI output by considering severity of harm, probability of occurrence, and detectability of the error before it causes harm |
| Risk Evaluation | Combined Risk (Influence & Consequence) | Integrate model influence and decision consequence to assign a risk tier (Low, Medium, Medium-High, High) |
| Risk Treatment | Oversight Mechanism | Implement human oversight based on risk tier |
| Monitoring & Review | Dynamic Oversight | Continuously monitor AI performance and context of use, adjusting the level or mode of oversight as the system or associated risks evolve |

**Table 1.** Mapping of risk evaluation with human oversight (Authors' proposal)

## 5. Illustrative risk assessments informing human oversight

To illustrate the proposed framework (Figure 5), including the relationship between model influence, decision consequence, and the corresponding level of human oversight, three representative scenarios are examined across the domains of finance, education, and healthcare. Each case demonstrates how AI systems can have different levels of risk, depending on the system's purpose, context of use, impact, and consequences. The risk assessment (ISO, 2018; ASME, 2018) framework provides a structured basis for assessing model risk by combining decision consequence (the potential impact if the model fails) with model influence (the extent to which the model informs or drives decisions). Further, these are aligned with human oversight mechanisms (EU, 2019). Each case therefore illustrates the application of distinct human oversight models—Human-in-Command (HIC), Human-in-the-Loop (HITL), and Human-on-the-Loop (HOTL)—as mechanisms for maintaining accountability, fairness, and regulatory compliance in AI-enabled decision-making contexts. As can be observed, model risk is distinct from technical flaws, it pertains to the potential for harm when decisions are made based on flawed or misleading outputs (Pandit & Rintamäki, 2024). Human oversight, through mechanisms like HIC, HITL, and HOTL, helps mitigate such risks by addressing challenges such as opacity, bias, and lifecycle limitations while promoting fairness, safety, and compliance (EU, 2019). Regardless of the nature of the risk, human oversight helps detect and correct errors, promote fairness, maintain trust, and ensure that AI systems align with regulatory requirements, operational goals, and societal values. Each of the scenarios presented below to showcase the framework's potential applicability is inspired by existing analyses and case studies.

**Scenario 1 (finance)**

AI/ML-based prediction for bank loan approvals (see, for example, Haque & Hassan, 2024; Sheikh et al., 2024; Uddin et al., 2023)

| Aspect | Details |
|---|---|
| **Context** | Banks employ AI/ML models to automate credit approval decisions. These models analyze applicant data, including income, credit history, loan purpose, to predict the likelihood of default and recommend whether to approve or deny a loan. |
| **Model Influence:** | **High.** The AI system outputs directly impact financial decisions, affecting individuals' access to credit. |
| **Decision Consequence** | **High**, as incorrect decisions can result in financial losses, regulatory penalties, reputational damage, and violations of fair-lending regulations. |
| **Evaluation** | Overall, the AI system risk is **High** (Figure 4). |
| **Human Oversight Mechanism** | **Human-in-Command (HIC) (Figure 5)**<br>Final decision authority rests with a loan officer or risk manager who reviews AI-generated recommendations, considers exceptional circumstances (e.g., a recent job change or temporary financial hardship), and either approves or overrides the decision. The human is not merely reviewing edge cases but actively controls when and how AI is deployed, and can pause its use if systemic bias is detected. |

**Scenario 2 (education)**

AI and student performance prediction (see, for example, Ahmed et al., 2025; Hoti et al., 2025; Johora et al., 2025; Karale et al., 2022; Tirumanadham et al., 2025)

| Aspect | Details |
|---|---|
| **Context** | AI/ML models analyze historical academic records to identify students at risk of dropping out or underperforming. The models also generate personalized study or career recommendations. |
| **Model Influence** | **High.** The AI system significantly impacts decisions related to intervention, including tutoring programs and career counselling sessions. Its predictions are central to institutional strategy for student success. |
| **Decision Consequence** | **Medium**, as incorrect predictions can lead to reputational damage for the institution, demotivation or stigmatization of students, and inequitable outcomes. Inaccurate career recommendations may also affect long-term employability and satisfaction. |
| **Evaluation** | Overall, the AI system risk is **Medium-High** (Figure 4). |
| **Human Oversight Mechanism** | **Human-in-the-Loop (HITL) (Figure 5)**<br>Academic counsellors and/or teachers review AI-generated risk profiles and recommendations before taking action. They discuss with students, validate the predictions using qualitative insights (e.g., recent personal issues affecting performance), and adjust interventions accordingly. |

**Scenario 3 (healthcare)**

Automated patient scheduling using AI (see, for example, Algarvio, 2025; Knight et al., 2023; Schmider, 2019)

| Aspect | Details |
|---|---|
| **Context** | An AI system automates patient appointment scheduling in a clinic or hospital. While errors may lead to delays, missed appointments, or minor patient dissatisfaction, they are unlikely to cause direct harm patient health. |
| **Model Influence:** | **Medium.** The AI system in this scenario does not make safety-critical decisions, as it is only involved in scheduling patients' appointments. |
| **Decision Consequence** | **Low**, as incorrect scheduling may inconvenience patients or staff, but it does not directly affect patient safety. |
| **Evaluation** | Overall, the AI system risk is **Low to Medium** (Figure 4). |
| **Human Oversight Mechanism** | **Human-on-the-Loop (HOTL) (Figure 5)** Humans monitor system outputs can intervene if anomalies occur, but daily operations largely proceed autonomously. |

## 5. Conclusion

The responsible implementation of high-risk AI systems requires a nuanced understanding of both their transformative potential and their inherent risks. Effective risk assessment frameworks must account for the specific context of use, the design intentions, and the emergent risks arising from AI errors, bias, or opacity. While HITL oversight can be valuable, applying it uniformly to all AI systems may be time-consuming and burdensome. Consistent with guidance from the EU AI Act, the level of human oversight should therefore be proportionate to risk, with oversight intensity determined by factors such as the severity, probability, and detectability of potential harms (European Commission, 2021). Integrating tailored human oversight strategies—whether Human-in-Command, Human-in-the-Loop, or Human-on-the-Loop—not only enhances transparency and accountability but also helps ensure that AI-driven decisions remain aligned with ethical, legal, and societal standards (Floridi et al., 2018). Ongoing adaptation of risk assessment methods, in accordance with evolving technologies, is essential to maintain trust, protect fundamental rights, and maximize the benefits of AI innovation. Ultimately, balancing opportunity and risk through robust assessment and governance enables organizations to harness AI systems safely and fairly, fostering innovation while minimizing potential harms.

Chu, Y., & Liu, P. (2023). Machines and humans in sacrificial moral dilemmas: Required similarly but judged differently? *Cognition*, 239, 105575. https://doi.org/10.1016/j.cognition.2023.105575

Choudhury, A., & Asan, O. (2020). Role of artificial intelligence in patient safety outcomes: Systematic literature review. *JMIR Medical Informatics*, 8(7), e18599. https://doi.org/10.2196/18599

Corrêa, N.K., Galvão, Santos, J.W., et al. (2023). Worldwide AI ethics: A review of 200 guidelines and recommendations for AI governance. *Patterns*, 4(10), 100857. https://doi.org/10.1016/j.patter.2023.100857

Crootof, R., Kaminski, M.E., & Nicholson Price II, W. (2023). Humans in the loop, *Vanderbilt Law Review*, 76(2), 429–510. https://scholarship.law.vanderbilt.edu/vlr/vol76/iss2/2

de Almeida, P.G.R., & dos Santos, C.D. (2025). Artificial intelligence governance: Understanding how public organizations implement it. *Government Information Quarterly*, 42(1), 102003. https://doi.org/10.1016/j.giq.2024.102003

DiFrancesco, J. (2025). The FDA's draft guidance, use of AI in regulatory decision-making for drug & biological products. *Avancer*, January 8. https://avancer.co/our-insights/articles/articles/harnessing-artificial-intelligence-in-drug-and-biological-product-development-a-strategic-overview-of-the-fdas-draft-guidance-for-stakeholders#:~:text=,the%20necessary%20depth%20of%20testing

Dubber, M.D., Pasquale, F., & Das, S. (Eds.) (2021). *The Oxford Handbook of Ethics of AI*. Oxford: Oxford University Press.

Elgin, C.Y., & Elgin, C. (2024). Ethical implications of AI-driven clinical decision support systems on healthcare resource allocation: A qualitative study of healthcare professionals' perspectives. *BMC Medical Ethics*, 25(1), 148. https://doi.org/10.1186/s12910-024-01151-8

Emami, Y., Almeida, L., Li, K., et al. (2024). Human-in-the-loop machine learning for safe and ethical autonomous vehicles: Principles, challenges, and opportunities. *arXiv*. https://doi.org/10.48550/arXiv.2408.12548

Enarsson, T., Enqvist, L., & Naarttijärvi, M. (2022). Approaching the human in the loop: Legal perspectives on hybrid human/algorithmic decision-making in three contexts. *Information & Communications Technology Law*, 31(1), 123–153. https://doi.org/10.1080/13600834.2021.1958860

EU. (2019). Ethics guidelines for trustworthy AI. https://digital-strategy.ec.europa.eu/en/library/ethics-guidelines-trustworthy-ai

EU. (2024). EU Artificial Intelligence Act. https://eur-lex.europa.eu/legal-content/EN/TXT/?uri=CELEX%3A32024R1689

Fabiano, N. (2025). AI Act and Large Language Models (LLMs): When critical issues and privacy impact require human and ethical oversight. *arXiv*. https://doi.org/10.48550/arXiv.2404.00600

Ferrara, M., Bertozzi, G., Di Fazio, N., et al. (2024). Risk management and patient safety in the artificial intelligence era: A systematic review. *Healthcare*, 12(5). https://doi.org/10.3390/healthcare12050549

Floridi, L., Cowls, J., Beltrametti, M., et al. (2018). *AI4People—An ethical framework for a good AI society: Opportunities, risks, principles, and recommendations*. Minds and Machines, 28(4), 689–707. https://doi.org/10.1007/s11023-018-9482-5

Fui-Hoon Nah, F., Zheng, R., Cai, J., et al. (2023). Generative AI and ChatGPT: Applications, challenges, and AI–human collaboration. *Journal of Information Technology Case and Application Research*, 25(3), 277–304. https://doi.org/10.1080/15228053.2023.2233814

Gardiner, B. (2024). How Silicon Valley is disrupting democracy. *MIT Technology Review*, December 13. https://www.technologyreview.com/2024/12/13/1108459/book-review-silicon-valley-democracy-techlash-rob-lalka-venture-alchemists-marietje-schaake-tech-coup/amp/

Gebru, T., & Torres, É.P. (2024). The TESCREAL bundle: Eugenics and the promise of utopia through artificial general intelligence. *First Monday*, 29(4). https://doi.org/10.5210/fm.v29i4.13636
15

\* \* \*